\documentclass[runningheads]{llncs}

\usepackage[T1]{fontenc}
\usepackage{amsmath,amssymb,amsfonts,bm}
\usepackage{mathtools}
\usepackage{graphicx}
\usepackage{booktabs}
\usepackage{siunitx}
\usepackage{xcolor}
\usepackage{url}
\usepackage[capitalise,noabbrev]{cleveref}

\begin{document}

\title{Uncertainty-Aware Multimodal Fusion for Oral Lesion Classification}

% \author{Anonymized Authors}
% \authorrunning{Anonymized Author et al.}
% \institute{Anonymized Affiliations \\
%     \email{email@anonymized.com}}

% For Final Submission
\author{%
Soujanya Hazra $^{\dagger}$ \inst{1}  \and 
Rupam Mukherjee $^{\dagger}$ \inst{2}
Rajkumar Daniel $^{\dagger}$ \inst{2} \and 
Shirin Dasgupta\inst{3} \and 
Subhamoy Mandal\inst{2}
\thanks{%
Corresponding author and $^{\dagger}$These authors contributed equally.}
}%
\authorrunning{Hazra et al.}
\institute{%
Department of Electrical Engineering, IIT Kharagpur, India \\
\and
School of Medical Science \& Technology, IIT Kharagpur, India\\
\email{ smandal@smst.iitkgp.ac.in}
\and
Dr. B.C. Roy Multispeciality Medical Research Centre, IIT Kharagpur, India
}

\maketitle
%%%%%%%%%%%%%%%%%%%%%%%%%
\begin{abstract}
Early detection of oral cancer and potentially malignant diseases is a major challenge in low-resource settings due to the scarcity of annotated data. We provide a unified approach for oral lesion classification that incorporates deep learning, spectral analysis, and demographic data. A pathologist-verified subset of oral cavity images was curated from a publicly available dataset. Oral cavity pictures were processed using a fine-tuned ConvNeXt-v2 network for deep embeddings before being translated into the hyperspectral domain using a reconstruction algorithm. Haemoglobin–sensitive, textural, and spectral descriptors were obtained from the reconstructed hyperspectral cubes and combined with demographic data. Multiple machine-learning models were evaluated using patient-specific validation. Finally, an incremental heuristic meta-learner (IHML) was developed that merged calibrated base classifiers via probabilistic feature stacking and uncertainty-aware abstraction of multimodal representations with patient-level smoothing. 
By decoupling evidence extraction from decision fusion, IHML stabilizes predictions in heterogeneous, small-sample medical datasets.
On an unseen test set, our proposed model achieved a macro F1 of 66.23\% and an overall accuracy of 64.56\%. The findings demonstrate that RGB-to-hyperspectral reconstruction and ensemble meta-learning improve diagnostic robustness in real-world oral lesion screening.
\end{abstract}
%%%%%%%%%%%%%%%%%%%%%%%%%%%%%%%%%%%%%%%%%%%%%%%%%%
\begin{keywords}
Multi-modal learning, oral lesion classification, heuristics, hyperspectral features, meta-classifier.
\end{keywords}
%%%%%%%%%%%%%%%%%%%%%%%%%%%%%%%%%%%%%%%%%%%%%%%%%%
\section{Introduction and Related Works}
\label{sec:introduction}
Oral cancer (OCA) and oral potentially malignant disorders (OPMD) are still significant causes of morbidity and mortality, particularly in South and Southeast Asia. Early detection is crucial, but clinical screening is often hindered by subjectivity and a shortage of professionals \cite{who2022oralhealth}. Artificial intelligence and computer vision, particularly deep convolutional models and transfer learning, have shown strong potential for improving lesion detection and risk assessment.  
Existing datasets, limited by small sample sizes, inconsistent collection methods, and poor clinical metadata, continue to affect the reliability and practical translation of AI-based oral screening systems. The recent study ~\cite{piyarathne2024oraloncology} aims to close this gap by enabling image-based classification enriched with contextual factors, such as age, gender, and behavioral risks. However, most existing methods rely solely on RGB images, overlooking spectral and textural information that could enhance lesion classification. Advances in image restoration networks have enhanced the recovery of fine structural details in impaired medical and photographic data.  The multi-stage progressive restoration model MPRNet~\cite{zamir2021mprnet} enhances texture, contrast, and lighting uniformity, improving visual quality while preserving diagnostically relevant cues for feature extraction.  In parallel, ConvNeXt-v2~\cite{liu2023convnextv2} offers high-quality embeddings with modern convolution efficiency.  
On the classification side, ensemble learning and meta-modeling techniques have demonstrated strong generalization in heterogeneous biomedical data~\cite{ke2017lightgbm}. Gradient-boosting and tree-based models effectively capture nonlinear dependencies, but combining multiple calibrated learners can further stabilize predictions and handle multimodal feature spaces.
However, existing multimodal oral lesion studies typically rely on either single-stage feature concatenation or end-to-end classifiers, which are sensitive to feature redundancy, dataset imbalance, and intra-patient variability. In contrast, our approach explicitly models predictive uncertainty and enforces patient-level consistency, both of which are critical for real-world screening scenarios with heterogeneous image quality.
Our key contributions are as follows:
(1) We present a multimodal ensemble process for four-class oral lesion classification using RGB deep embeddings, hyperspectral reconstruction, customized spectral–textural descriptors, and demographic metadata.
(2) To ensure diagnostic reliability, we curate a pathologist-verified subset of the dataset, process each image with a fine-tuned encoder, and rebuild 31-band hyperspectral cubes to generate physiologically significant features.
(3) For reliable and uncertainty-aware predictions, we build an IHML that combines calibrated classifier probabilities via probabilistic stacking and patient-level posterior smoothing.
(4) We demonstrate high and consistent performance on a patient-wise unseen test split, demonstrating that deep morphological cues, spectral indices, and clinical priors increase real-world oral lesion screening diagnostic robustness.
%%%%%%%%%%%%%%%%%%%%%%%%%%%%%%%%%%%%%%%%%%%%%%%%%
\section{Methodology}
\label{sec:methodology}
\begin{figure}
    \centering
    \includegraphics[width=1\textwidth]{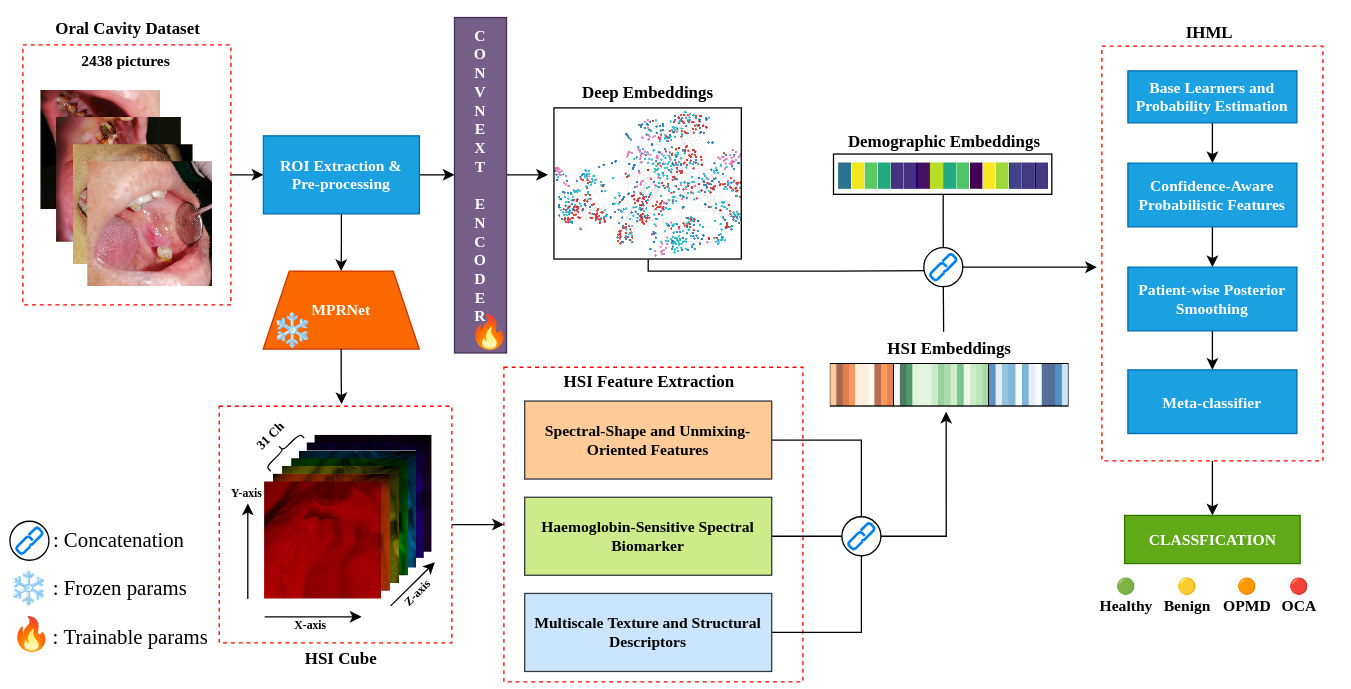}
    \caption{Proposed multimodal pipeline uses RGB deep embeddings, hyperspectral characteristics, and demographic data to predict oral lesions in patients using IHML.
    }
    \label{fig:pipeline}
\end{figure}

The proposed pipeline, illustrated in Fig.~\ref{fig:pipeline}, combines deep morphological embeddings, handcrafted spectral features, and patient-level clinical data into a unified multimodal representation for oral lesion classification.
%%%%%%%%%%%%%%%%%%%%%%%%%%%%%%%%%%%%%%%%%%%%%%%%%
%%%%%%%%%%%%%%%%%%%%%%%%%%%%%%%%%%%%%%%%%%%%%%%%%
\subsection{Generating Hyperspectral Images from RGB Images}
\begin{figure}[htbp]
    \centering

    \begin{minipage}{0.26\textwidth}
        \centering
        \includegraphics[width=\linewidth]{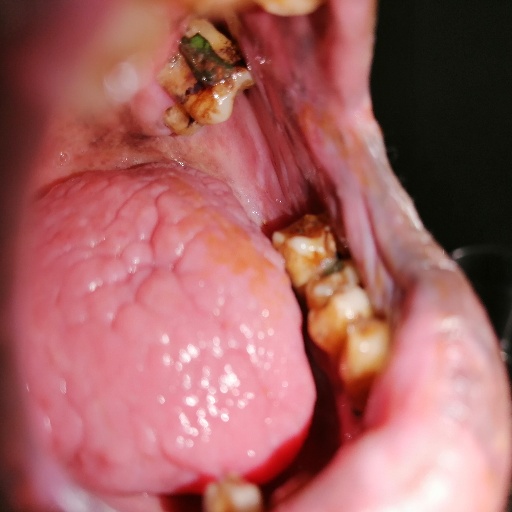}
        \\[-0.5em]
        (a)
    \end{minipage}
    \hfill
    \begin{minipage}{0.32\textwidth}
        \centering
        \includegraphics[width=\linewidth]{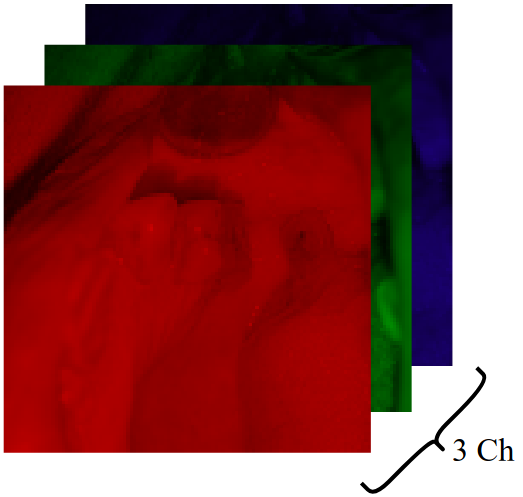}
        \\[-0.5em]
        (b)
    \end{minipage}
    \hfill
    \begin{minipage}{0.35\textwidth}
        \centering
        \includegraphics[width=\linewidth]{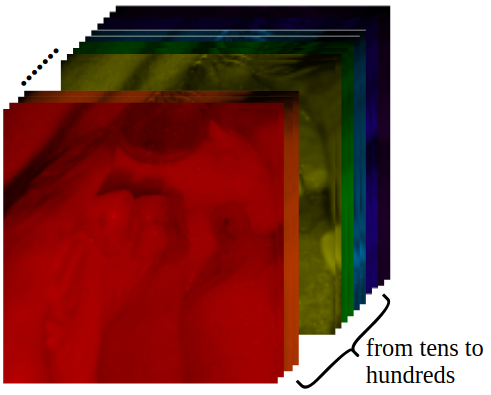}
        \\[-0.5em]
        (c)
    \end{minipage}

    \caption{Examples of RGB and hyperspectral images. (a) displays the subject. (b) depicts the channel configuration with a general camera, while (c) shows the channel structure with a hyperspectral camera.}
    \label{fig:rgb_hsi}
\end{figure}

Hyperspectral photos capture reflectance across tens to hundreds of small spectral bands, usually across the visible spectrum, unlike RGB images, which use only three broad channels for red, green, and blue wavelengths.
True hyperspectral oral images require expensive, complex equipment, making them unsuitable for large-scale or low-resource screening. 
Research on producing hyperspectral images from RGB images began in the early 2000s~\cite{Parmar2008recon}, and in recent years, CNNs and other machine learning approaches have been widely applied~\cite{Stiebel2018recon,zhao2020hierarchical}.
In this work, we fine-tuned the Multi-Stage Progressive Image Restoration Network (MPRNet)~\cite{zamir2021mprnet} to recover hyperspectral representations from RGB oral lesion images. The network creates a hyperspectral cube $H \in \mathbb{R}^{512 \times 512 \times 31}$ from an RGB region of interest $I \in \mathbb{R}^{512 \times 512 \times 3}$, sampling spectral bands at 10-nm intervals from 400-700 nm. The RGB image is extended along the spectral dimension using the reconstructed cube, preserving the image-plane spatial structure.
Fig.~\ref{fig:rgb_hsi} depicts the conceptual difference between RGB and hyperspectral representations. The spatial dimensions $(x, y)$ correspond to the image plane, but the spectral dimension $(z)$ stores wavelength-dependent reflectance differences across many channels, which are then used for feature extraction. 
% In particular, MPRNet~\cite{zamir2021mprnet}
% uses the peak signal-to-noise ratio (PSNR) as an index to evaluate the accuracy of the generated hyperspectral image.
% A PSNR of 33.50 dB indicates consistent reconstruction quality suitable for downstream feature extraction.
The reconstructed hyperspectral representations provide spectrally enriched cues for wavelength-localized feature extraction from RGB images in resource-constrained settings, rather than replacing physically acquired HSI.
%%%%%%%%%%%%%%%%%%%%%%%%%%%%%%%%%%%%%%%%%%%%%%%%%
%%%%%%%%%%%%%%%%%%%%%%%%%%%%%%%%%%%%%%%%%%%%%%%%%
\subsection{Deep Representation Extraction}
Following hyperspectral reconstruction, we extract various deep and handcrafted representations to capture global morphological patterns of oral lesions and fine-grained spectral–textural features.
Deep morphological features are recovered from the RGB region of interest (ROI) using ConvNeXt-v2~\cite{liu2023convnextv2}. The network is initialized with ImageNet-pretrained weights and fine-tuned on the curated oral lesion dataset to match domain-specific visual patterns. After training, the classification head is removed, and the output of the final global average pooling layer is used as a compact deep embedding. This computationally efficient approach encodes high-level structural information, color gradients, and contextual lesion morphology.
Each ROI's reconstructed hyperspectral cube is kept for downstream feature computation. 
%%%%%%%%%%%%%%%%%%%%%%%%%%%%%%%%%%%%%%%%%%%%%%%%%
\subsection{Feature Set for Lesion Separation}
We extracted three complementary families of features that provide a compact yet physiologically grounded representation for the lesion.
%%%%%%%%%%%%%%%%%%%%%%%%%
\subsubsection{Haemoglobin–Sensitive Spectral Biomarkers:}
Oxy- and deoxy-haemoglobin exhibit characteristic absorption behaviour around 545-575\,nm. We extract
multiple ratiometric features that emphasize relative attenuation in these bands. These include the well-established reflectance ratio $R_{545}/R_{575}$, the normalized difference index 
$\mathrm{NDI} = (R_{545}-R_{575})/(R_{545}+R_{575})$~\cite{guo2024periodontal}, a pseudo-Hb contrast using 
$(R_{560}-R_{600})$, and short-range spectral scatter slopes estimated by local linear fits across narrow spectral windows.  
%%%%%%%%%%%%%%%%%%%%%%%%%
\subsubsection{Multiscale Texture and Structural Descriptors:}
We extracted texture features that capture the spatial patterns present in oral lesions. Grey-level co-occurrence matrix (GLCM) statistics describe second-order textures such as contrast and homogeneity. Local Binary Patterns (LBP) capture fine surface details and local intensity changes. Gabor filters measure texture at multiple scales and orientations. SIFT descriptors detect~\cite{goswami2025multimodal} and summarize key local structures within the lesion regions.  
%%%%%%%%%%%%%%%%%%%%%%%%%
\subsubsection{Spectral-Shape and Unmixing-Oriented Features:}
We also extract features that describe the overall shape of the reflectance spectrum. These include band-wise maxima and minima, the wavelengths at which they occur, peak-to-valley differences, and simple slope and curvature measures. Such features reflect changes in chromophore composition, including keratin, melanin, and oxygenated blood~\cite{jacques2013optical}. These descriptors complement haemoglobin-sensitive ratios by capturing global spectral trends that are less sensitive to absolute reflectance magnitude and illumination variability.
%%%%%%%%%%%%%%%%%%%%%%%%%%%%%%%%%%%%%%%%%%%%%%%%%
\subsection{Multimodal Feature Fusion}
Deep ConvNeXt-v2 embeddings and handcrafted feature sets capture complementary aspects of lesion appearance. The deep representation encodes global morphology, color gradients, and high-level contextual structure. At the same time, the handmade descriptors give fine-grained spectral variations, local texture patterns, and clinically relevant demographic risk factors.
To construct a unified representation, all feature modalities were concatenated into a single multimodal vector:
\[
\mathbf{x} = 
\mathbf{z}_{\mathrm{deep}}
\,\Vert\,
\mathbf{z}_{\mathrm{hae}}
\,\Vert\,
\mathbf{z}_{\mathrm{tex}}
\,\Vert\,
\mathbf{z}_{\mathrm{spec}}
\,\Vert\,
\mathbf{z}_{\mathrm{demo}},
\]
where 
$\mathbf{z}_{\mathrm{deep}} \in \mathbb{R}^{768}$, 
$\mathbf{z}_{\mathrm{hae}} \in \mathbb{R}^{46}$,
$\mathbf{z}_{\mathrm{tex}} \in \mathbb{R}^{58}$,
$\mathbf{z}_{\mathrm{spec}} \in \mathbb{R}^{31}$,
and $\mathbf{z}_{\mathrm{demo}} \in \mathbb{R}^{5}$ 
represents the ConvNeXt-v2, haemoglobin-sensitive, texture, spectral-style, and demographic features, respectively. 
Potential feature collinearity is mitigated through architectural design rather than linear compression. 
Specifically, features are normalized independently within each modality to avoid scale-induced dominance, and the subsequent IHML uses calibrated probabilistic outputs rather than raw feature activations. 
Tree-based base learners use hierarchical splits to suppress redundant dimensions, whereas the meta-learning stage aggregates uncertainty-aware posterior statistics, compressing correlated evidence into low-variance confidence measures. 
%%%%%%%%%%%%%%%%%%%%%%%%%%%%%%%%%%%%%%%%%%%%%%%%%
\subsection{Incremental Heuristic Meta-Learner}
It incorporates the decisions from multiple heterogeneous classifiers, extracts uncertainty-aware meta-features, and uses iterative posterior smoothing to ensure patient-level prediction consistency. This architecture enables robust multiclass classification despite varying image quality and significant intra-patient correlation.
%%%%%%%%%%%%%%%%%%%%%%%%%
% \subsubsection{Base Learners and Probability Estimation:}
For each multimodal feature vector $\mathbf{x}$, we train four calibrated base models:
LightGBM, extra trees, gradient boosting, and isotonic-calibrated logistic regression.
Each model $g_m$ generates a probability distribution for the four lesion classes:% \[
$
\mathbf{p}^{(m)} = g_m(\mathbf{x}) \in [0,1]^{4}.
% \]
$
These probability vectors constitute the first layer of the IHML stack.
%%%%%%%%%%%%%%%%%%%%%%%%%
% \subsubsection{Confidence-Aware Probabilistic Features:}

We use scalar confidence statistics from $\mathbf{p}^{(m)}$ to assess the reliability of each base model.  These include the highest class confidence, the top-two class margin, and the Shannon entropy.  We denote this transformation as $\psi(\cdot)$:
% \[
$
\mathbf{c}^{(m)} = \psi\big(\mathbf{p}^{(m)}\big),
% \]
$ where $\mathbf{c}^{(m)}$ is a low-dimensional vector encoding uncertainty and inter-class separation.  
The meta-feature vector is created by concatenating all base probabilities and confidence features.
% $
\setlength{\abovedisplayskip}{2pt}
\setlength{\belowdisplayskip}{2pt}
\setlength{\abovedisplayshortskip}{0pt}
\setlength{\belowdisplayshortskip}{0pt}
\begin{equation}
\mathbf{h}
=
\Phi\!\left(
\mathbf{p}^{(1)},\dots,\mathbf{p}^{(M)},
\mathbf{c}^{(1)},\dots,\mathbf{c}^{(M)}
\right)
\end{equation}
where $\Phi(\cdot)$ stacks all components into a single representation.

%%%%%%%%%%%%%%%%%%%%%%%%%
% \subsubsection{Patient-wise Posterior Smoothing:}
Because multiple images come from the same patient, the classifier's outputs may exhibit unexpected intra-subject variation. IHML mitigates this effect through a patient-level refining stage.  We iteratively update the probability vector for sample $i$ in patient group $g(i)$ as follows:
\begin{equation}
\mathbf{p}^{(t+1)}_{i}
=
(1-\alpha)\,\mathbf{p}^{(t)}_{i}
+
\alpha\,\overline{\mathbf{p}}^{\,(t)}_{g(i)},
\end{equation}
where $\overline{\mathbf{p}}_{g(i)}$ is the mean probability of all samples from the same patient, and $\alpha$ controls the influence of the group prior. This technique stabilizes predictions and reflects the clinical fact that a patient's lesion category remains constant across adjacent images.

%%%%%%%%%%%%%%%%%%%%%%%%%
% \subsubsection{Meta-classifier:}
A multinomial logistic regression model trained on $\mathbf{h}$ determines the final decision:
\begin{equation}
\hat{y}
=
\arg\max_{c}
\;
\sigma\!\left(
\mathbf{W}\mathbf{h} + \mathbf{b}
\right)_c,
\end{equation}
where $\sigma$ denotes the softmax function. 
%%%%%%%%%%%%%%%%%%%%%%%%%
IHML improves prediction stability by combining diverse learners in probability space, modeling uncertainty, and enforcing patient-level consistency. Unlike end-to-end tabular models or conventional stacking, it separates evidence extraction from decision fusion and refines predictions through group-consistent Bayesian smoothing. This reduces sensitivity to feature redundancy and class imbalance, improving robustness when training data is limited.
%%%%%%%%%%%%%%%%%%%%%%%%%%%%%%%%%%%%%%%%%%%%%%%%%
\section{Experiments and Results}
\subsection{Dataset and Experimental Setup}
We used an openly available oral lesion dataset~\cite{piyarathne2024oraloncology} containing 3,000 white-light oral cavity images from 714 patients, labeled as healthy, benign, OPMD, or OCA. The dataset also includes polygonal lesion or mouth-cavity annotations and patient metadata, including age, gender, smoking, alcohol use, and betel chewing.
All images were reviewed by an in-house oral pathologist, and diagnostically unreliable images were excluded due to artifacts such as defocus, motion blur, poor illumination, overexposure, shadows, or poor lesion visibility. The remaining images were cropped to the lesion region of interest and resized to \(512\times512\) pixels. The final curated cohort contained 2,438 images from 653 patients.
The cohort was naturally imbalanced, reflecting real-world screening settings: OPMD accounted for 46.5\% of images, followed by benign lesions (24.9\%), healthy samples (24.3\%), and OCA (4.3\%). All experiments used stratified subject-wise splitting. Specifically, 15\% of patients were held out as an unseen test set, while the remaining 85\% formed the development set. Model selection was performed using 5-fold cross-validation on the development set, and final performance was reported on the held-out patient test set.
%%%%%%%%%%%%%%%%%%%%%%%%%
\subsection{Performance Comparison on Models}
\begin{figure}[t]
    \centering
    \includegraphics[width=0.90\textwidth]{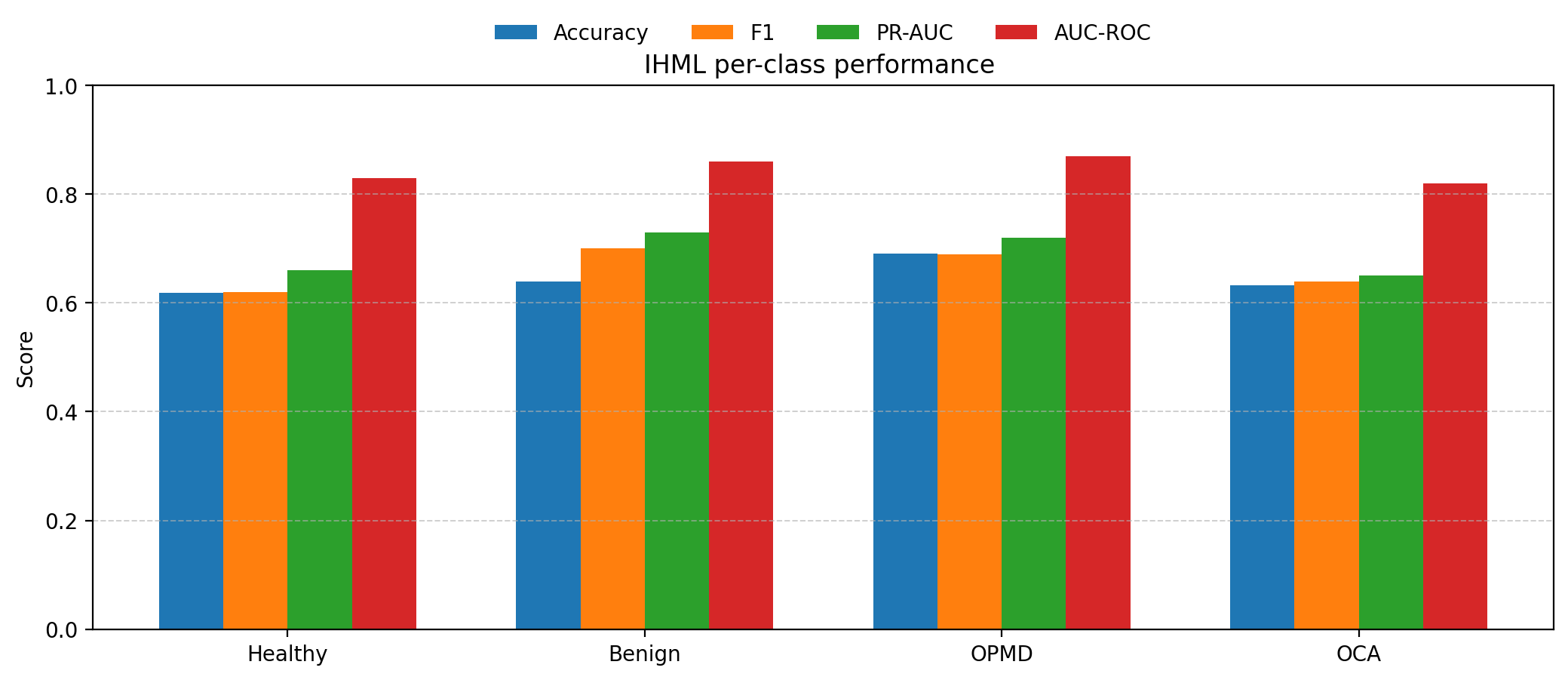}
    \caption{Per-class performance of the proposed model on the held-out test set.}
    \label{fig:per_class_perf}
\end{figure}

We evaluated the fused multimodal representation using classical baselines, including logistic regression, random forest, SVM, XGBoost, and LightGBM, as well as recent deep tabular models such as TabICL~\cite{qu2025tabicl}, T2G-Former~\cite{yan2023t2g}, TabTransformer~\cite{huang2020tabtransformer}, and DANet~\cite{chen2022danets}.
Table~\ref{tab:mainresults} shows that the proposed IHML outperforms both conventional and deep tabular baselines across all evaluation metrics. This improvement reflects the benefit of calibrated heterogeneous learners, uncertainty-aware probabilistic fusion, and patient-level posterior smoothing. Fig.~\ref{fig:per_class_perf} further shows that IHML maintains balanced per-class performance across healthy, benign, OPMD, and OCA categories despite class imbalance.
%%%%%%%%%%%%%%%%%%%%%
\begin{table}
\centering
\caption{Comparison results on the patient-wise held-out test set.}
\label{tab:mainresults}
\begin{tabular}{l|c|c|c|c}
\hline
\textbf{Model} & \textbf{Macro F1} & \textbf{Accuracy} & \textbf{PR-AUC} & \textbf{AUC-ROC} \\
\hline
Logistic Regression & 44.03 & 47.10 & 45.86 & 71.70 \\
Random Forest & 58.08 & 58.82 & 65.24 & 81.03 \\
SVM & 54.20 & 54.41 & 63.45 & 81.47 \\
XGBoost & 55.44 & 57.35 & 66.65 & 82.54 \\
LightGBM & 61.27 & 59.81 & 67.51 & 82.89 \\
\hline
TabICL & 57.14 & 60.29 & 62.04 & 78.99 \\
T2G-Former & 44.51 & 49.35 & 48.57 & 74.98 \\
TabTransformer & 47.21 & 47.79 & 50.38 & 71.42 \\
DANet & 57.20 & 61.76 & 66.03 & 83.09 \\
\hline
\textbf{Proposed} & \textbf{66.23} & \textbf{64.56} & \textbf{69.10} & \textbf{84.45} \\
\hline
\end{tabular}
\end{table}
%%%%%%%%%%%%%%%%%%%%%
%%%%%%%%%%%%%%%%%%%%%
\subsection{Ablation Study}
We conducted a stepwise ablation study, incrementally adding feature groups to the baseline and measuring the contribution of each feature family within the IHML framework. Table~\ref{tab:ablation} reports five configurations (M1-M5), with checkmarks indicating the presence of a feature block and crosses representing its removal. Model M1 only embeds ConvNeXt. M2 includes demographic characteristics indicating patient clinical priors. M3 uses haemoglobin–sensitive descriptors from the reconstructed spectral cube. M4 adds texture information, while M5 has all spectrum descriptors. The steady improvement proves that each feature family provides supplementary information. 
%%%%%%%%%%%%%%%%%%%%%
\begin{table}
\centering
\caption{Ablation study evaluating the contribution of individual feature groups within the IHML framework. \textbf{D} denotes demographic features, \textbf{H} denotes haemoglobin-sensitive spectral biomarkers, \textbf{T} denotes multiscale texture and structural descriptors, and \textbf{S} denotes spectral-shape and unmixing-oriented features.}
\label{tab:ablation}
\begin{tabular}{l|cccc|c|c|c|c}
\hline
\textbf{Model} & \textbf{D} & \textbf{H} & \textbf{T} & \textbf{S} & \textbf{Macro F1} & \textbf{Accuracy} & \textbf{PR-AUC} & \textbf{AUC-ROC} \\
\hline
M1 & \(\times\) & \(\times\) & \(\times\) & \(\times\) & 54.97 & 52.52 & 64.44 & 83.86 \\
M2 & \(\checkmark\) & \(\times\) & \(\times\) & \(\times\) & 60.08 & 58.27 & 68.12 & 81.03 \\
M3 & \(\checkmark\) & \(\checkmark\) & \(\times\) & \(\times\) & 63.19 & 62.24 & 63.38 & 82.84 \\
M4 & \(\checkmark\) & \(\checkmark\) & \(\checkmark\) & \(\times\) & 64.13 & 62.55 & 67.07 & 82.27 \\
\textbf{M5} & \(\checkmark\) & \(\checkmark\) & \(\checkmark\) & \(\checkmark\) & \textbf{66.23} & \textbf{64.56} & \textbf{69.10} & \textbf{84.45} \\
\hline
\end{tabular}
\end{table}
%%%%%%%%%%%%%%%%%%%%%%%%%%%%%%%%%%%%%%%%%%%%%%%%%
\section{Conclusion}
\label{sec:conc}
We presented a multimodal fusion framework for oral lesion classification that integrates deep visual embeddings, spectrally enriched descriptors, and clinical priors. Our method uses deep morphological embeddings, haemoglobin-sensitive spectral indications, texture descriptors, and demographic data.  Each modality captures lesion appearance and clinical context uniquely.
RGB pictures were converted to 31-band hyperspectral images.  These extracted spectral-style biomarkers are not available in RGB imaging.  We created tiny texture and spectral-shape characteristics to capture local structure and wavelength-dependent variations.
We proposed integrating various features into IHML. It combines calibrated base learners, uncertainty-aware meta-features, and patient-level smoothing. This strategy improves stability and consistency in patient forecasts.
Our strategy was tested on an unobserved patient split. IHML has a macro F1-score of 66.23\% and an AUC-ROC of 84.45\%.  Our findings beat all baseline models.  Each feature block yielded considerable gains in the ablation research. Our findings demonstrate that incorporating deep, spectral, and clinical information enhances oral lesion categorization. 
Future work will explore end-to-end multimodal attention architectures with explicit uncertainty modeling, as well as validation on physically acquired hyperspectral datasets.
For better clinical application, we will investigate end-to-end fusion models, transformer-based attention mechanisms, and real-world hyperspectral imaging.

%%%%%%%%%%%%%%%%%%%%%%%%%
\subsubsection{Disclosure of Interests} The authors have no competing interests to declare
that are relevant to the content of this article.
%%%%%%%%%%%%%%%%%%%%%%%%%
% \newpage
\bibliographystyle{splncs04}
\bibliography{references}

\end{document}